\documentclass[notoc]{JHEP}
\usepackage{epsfig} 
\input{axodraw.sty}

\newbox\mybox
\newcommand\fverb{\setbox\mybox=\hbox\bgroup\verb}
\newcommand\fverbdo{\egroup\medskip\noindent\fbox{\unhbox\mybox}\ }
\newcommand\fverbit{\egroup\item[\fbox{\unhbox\mybox}]}
\font\beeg=cmr17 scaled 1600            

\newcommand\init[1]{\setbox\mybox=\hbox{{\beeg #1}~}%
                   \noindent\global\hangindent=\wd\mybox\global\hangafter-2%
                   \sc\smash{\llap {\lower 13.2pt \box\mybox}}}

\newcommand{\lwig}{\mbox{\,\raisebox{.3ex}
    {$<$}$\!\!\!\!\!$\raisebox{-.9ex}{$\sim$}\,}}
\newcommand{\gwig}{\mbox{\,\raisebox{.3ex}
    {$>$}$\!\!\!\!\!$\raisebox{-.9ex}{$\sim$}}\,}

\title{Developments in Deep-inelastic Structure Function Calculations}

\author{S. Moch$^a\,$\thanks{Talk at the EPS HEP 2001 conference, July
    12 - 18, 2001, Budapest (Hungary)} , J.A.M. Vermaseren$^b$ and M. Zhou$^b$ \\
$^a$Institut f{\"u}r Theoretische Teilchenphysik, 
  Universit{\"a}t Karlsruhe, \\ ~76128 Karlsruhe, Germany \\[1ex]
$^b${NIKHEF Theory Group, 
  Kruislaan 409, 1098 SJ Amsterdam, The Netherlands \\
}}

\conference{EPS HEP 2001, Budapest, Hungary}

\abstract{We review recent developments in the calculation of 
deep-inelastic structure functions to next-to-next-to leading 
order in perturbative QCD. 
We discuss the impact of these corrections on the determination 
of the strong coupling $\alpha_s$ and the parton distributions.
}

\preprint{NIKHEF-2001-008\\
          TTP-01-18}

\begin{document} 

{\init Structure} functions in inclusive deep-inelastic scattering 
offer the possibility for extremely precise determinations 
of the strong coupling $\alpha_s$ and the parton distribution functions.
The high statistical accuracy of the present experimental
measurements, and the data expected from the electron-proton collider 
HERA after the luminosity upgrade, demand analyses in perturbative QCD
beyond the standard next-to leading order (NLO) corrections.
In order to match the experimental precision, it is therefore necessary 
to calculate higher order perturbative QCD corrections for the 
structure functions $F_2,F_3$ and $F_L$, 
in particular the complete next-to-next-to leading order (NNLO) corrections.
This information is not fully available yet. Some time ago, the two-loop coefficient 
functions of $F_2,F_3$ and $F_L$ have been
calculated~\cite{vanNeerven:1991nn}, and more recently, they have been
completely checked~\cite{Moch:1999eb}. 
However, for the three-loop anomalous dimensions $\gamma_{\rm pp}^{(2)}$, 
only partial results are available thus far. These include a finite
number of fixed Mellin
moments~\cite{Larin:1991zw,Larin:1991tj,Larin:1997wd,Retey:2000nq}, 
both for $F_2$ and $F_3$, the large $n_f$-limit~\cite{Gracey:1994nn} 
of $\gamma_{\rm qq}^{(2)}$ and $\gamma_{\rm gg}^{(2)}$, in the latter case 
only the coefficient of the colour factor $n_f^2 C_A$, and several terms relevant in the
small-$x$ limit~\cite{Catani:1994sq}.

\section{Prospects}

To explore the prospects of a complete NNLO analysis in this situation, 
it has been a fruitful strategy to combine all theoretical 
information presently available together with reasonable assumptions 
about the parton distributions or the functional form of 
the three-loop splitting functions $P_{\rm pp}^{(2)}(x)$. 
These investigations resulted in approximate 
expressions~\cite{vanNeerven:1999ca}
for $P_{\rm pp}^{(2)}(x)$ with negligible residual uncertainties for $x \gwig 10^{-2}$. 
Subsequent studies of the NNLO evolution of parton distributions 
revealed a much reduced scale dependence in comparison to standard NLO analyses.
Together with recent developments to account for correlated experimental
errors~\cite{Botje:1999dj} in analyses and a rigorous
statistical treament~\cite{Giele:1998gw} of
parton distributions, these efforts seem to allow for quantitative 
estimates of the parton distribution uncertainties. 
Let us emphasize, that this task is of basic importance for precise 
luminosity measurements via W- and Z-boson production in upcoming
experiments at hadron colliders, and therefore essential for future
searches of the Higgs particle or effects of new physics.

The precision determination of the running coupling $\alpha_s$ based
on the available partial information about NNLO QCD corrections has also been 
performed.
The CCFR data for $xF_3$ from $\nu
N$-scattering~\cite{Seligman:1997mc} 
has been analyzed~\cite{Kataev:1997nc} using the fixed Mellin
moments~\cite{Larin:1991zw,Larin:1991tj,Larin:1997wd,Retey:2000nq} for $F_3$.
The SLAC and HERA data from 
$eP$-scattering~\cite{Benvenuti:1989rh} 
have been averaged with Bernstein polynomials to
extract $\alpha_s$ in a direct fit of Mellin moments to experimental 
data~\cite{Santiago:1999pr}.
The analyses agree within their errors and, more importantly,
indicate that an absolute error for the strong coupling $\Delta
\alpha_s \lwig 1 \%$ is possible.
In addition, the effect of higher twist contributions, i.e. terms suppressed as
$1/Q^2$, have been studied~\cite{Kataev:1997nc}
and the NNLO evolution of parton distributions has even been used to put
bounds on squark and gluino masses~\cite{Santiago:1999pr}.
Finally, it is worth mentioning 
that the knowledge of a number of fixed Mellin moments of the
three-loop coefficient functions~\cite{Larin:1991zw,Larin:1991tj,Larin:1997wd,Retey:2000nq} 
enables analyses even beyond NNLO and allows for 
estimates of the effect of the next-to-NNLO corrections.

In summary, with the complete NNLO QCD analyses for the structure functions 
$F_2,F_3$ and $F_L$ a new level of precision is reached in comparing theoretical
predictions to experimental data. Moreover, the chance arises to address new questions 
which could not be studied before.

\section{Progress}

Progress towards the calculation of the three-loop
anomalous dimensions and the coefficient functions crucially relies 
on the ability to perform all necessary loop integrations. A promising 
approach~\cite{Gonzalez-Arroyo:1979df,Kazakov:1988jk} which allows to calculate the integrals 
in Mellin space analytically as a general function of $N$, 
is based on the optical theorem and the operator product expansion (OPE).
The parameters of the OPE are directly related to the 
Mellin moments of the structure functions. 
For $F_{2}$ we can write
\begin{eqnarray}
\label{eq:F2mellin}
\displaystyle
F_{2}^N(Q^2)\!&=&\!     
\int\limits_0^1 dx\, x^{N-2} F_{2}(x,Q^2) \\
\!&=&\! 
\nonumber
\sum\limits_{j=\alpha,{\rm{q, g}}}
C_{2,j}^{N}\left(\frac{Q^2}{\mu^2},\alpha_s\right)
A_{{\rm{P}},N}^j\left(\mu^2\right)\, ,
\end{eqnarray}
and similar relations define $F_{3}^N$ and $F_{L}^N$.
Here, $C_{2,j}^{N}$ denote the coefficient functions and 
$A_{{\rm{P}},N}^j$ the spin averaged hadronic matrix elements 
of singlet operators $O^{\rm q}$, $O^{\rm g}$ and
non-singlet operators $O^{\alpha}$, $\alpha = 1,2,\dots,(n_f^2-1),$ 
of leading twist. 
The coefficient functions and the renormalized operator matrix elements 
in eq.(\ref{eq:F2mellin}) both satisfy renormalization group equations.
Due to current conservation they are governed by the same anomalous
dimensions $\gamma_{jk}$, which determine the scale evolution of deep-inelastic
structure functions, 
\begin{eqnarray}
\label{callanA}
{\lefteqn{
\sum_{k = \alpha,{\rm{q,g}}}
\Bigl[ \left\{ \mu^2 \frac{\partial}{\partial \mu^2} + \beta(\alpha_s(\mu^2))
 \frac{\partial}{\partial \alpha_s(\mu^2)} \right\} \delta_{jk} 
}} \nonumber \\
& & + \gamma_{jk}(\alpha_s(\mu^2))
\Bigr]
A_{{\rm{P}},N}^j\left(\mu^2\right)
=   0 \, ,\\[1ex]
\label{callanC}
{\lefteqn{
\sum_{k = \alpha,{\rm{q,g}}}
\Bigl[ \left\{ \mu^2 \frac{\partial}{\partial \mu^2} + \beta(\alpha_s(\mu^2))
 \frac{\partial}{\partial \alpha_s(\mu^2)} \right\} \delta_{jk} 
}} \nonumber \\
& & - \gamma_{jk}(\alpha_s(\mu^2))
\Bigr]
C^N_{2,k} \left(\frac{Q^2}{\mu^2},\alpha_s(\mu^2) \right) 
=   0 \, ,\quad\quad
\end{eqnarray} 
where $j = \alpha,{\rm{q,g}}$ and $\beta$ represents the
$\beta$-function of QCD.

The task is the calculation of the coefficient functions $C_{2,j}^{N}$ and the 
anomalous dimensions $\gamma_{jk}$ in dimensionally
regulated perturbation theory.
In practice, at a given order in perturbation theory, 
this amounts to calculating the $N$-th moment of all
contributing four-point diagrams with external partons of 
momentum $p, p^2 = 0$ and photons of momentum $q$, 
which is precisely the coefficient of $(p \cdot q/q^2)^N.$

In order to do so, it is useful to set up a hierarchy among all diagrams 
depending on the number of $p$-dependent propagators.
We define basic building blocks (BBB) as diagrams in which the
parton momentum $p$ flows only through a single line in the diagram. 
Composite building blocks (CBB) are all diagrams with more than one $p$-dependent propagator.
At the three-loop level, there are 10 BBB's, and 32 CBB's respectivley, which correspond to 
genuine three-loop topologies of the ladder, benz or non-planar type.
In addition, there exist numerous BBB's and CBB's at three loops with an effective
two-loop topology and a self-energy insertion in one line. 

For the BBB's, the single $p$-dependent propagator, 
say $1/(p-l)\cdot(p-l)$ in a loop with momentum $l$, can be expanded 
into a geometrical sum using $p^2 = 0$. Then, scaling arguments require 
the final answer for the $N$-th moment to be proportional to the 
coefficient $(2\,p\cdot l/l^2)^N$. 
Thus, one is left with two-point functions with symbolic 
powers of scalar products in the numerator and denominator. 
For these objects, one sets up a reduction scheme, that relates the
BBB under consideration to simpler diagrams, where certain
lines are eliminated, such that the topology simplifies.
For the CBB, a straightforward expansion of the $p$-dependent
propagators leads to multiple nested sums, which in general are very difficult to 
evaluate. 
Hence, one has to seek a reduction scheme, that maps a given CBB 
onto BBB's. This is achieved, if one can remove a $p$-dependent
propagator. If one can get rid of a $p$-independent propagator,
usually the topology simplifies.

For the BBB's and the CBB's alike the reduction schemes are determined 
with the help of integration-by-parts 
identities~\cite{'tHooft:1972fi,Chetyrkin:1981qh} and scaling 
identities~\cite{Moch:1999eb}.
The reduction identities often involve explicitly the parameter
$N$ of the Mellin moment and sometimes one has to set up difference
equations in $N$ for the $N$-th moment $F(N)$ of a diagram ,
\begin{eqnarray}
a_0(N) F(N) + a_1(N) F(N-1) + \\ 
\dots + a_n(N) F(N-n) + G(N) &=& 0 \, , \nonumber
  \label{diffeq}
\end{eqnarray}
where $G(N)$ denotes the $N$-th Mellin moment of simpler diagrams. 
In the reduction schemes for three-loop diagrams, 
we have encountered difference equations up to the third order.
First order difference equations can be solved at the cost of one sum 
over $\Gamma$-functions in dimensional regularization,
$D=4-2\epsilon$. The $\Gamma$-functions can be expanded in $\epsilon$
and the sum can be solved to any order in $\epsilon$ in terms of
harmonic sums~\cite{Vermaseren:1998uu,Blumlein:1998if}.
Higher order difference equations could be solved constructively. 
In general, the approach to calculate Mellin moments of 
structure functions relies on particular mathematical 
concepts~\cite{Vermaseren:2000we}, 
such as harmonic sums~\cite{Vermaseren:1998uu,Blumlein:1998if} 
and related sums~\cite{Borwein:1996yq}. Subsequently, 
the inverse Mellin transformation to $x$-space requires harmonic 
polylogarithms~\cite{Remiddi:1999ew}, or more general, 
multiple polylogarithms~\cite{MR2000c:11108}.
Difference equations for the evaluation of Feynman diagrams, although not in the context
of Mellin moments, have recently also been 
studied by other authors~\cite{Tarasov:2000sf,Laporta:2000dc}.

Let us present an example of a diagram of the non-planar type, 
that gives rise to a third order difference equation,
\begin{eqnarray}
  \label{no22bbbint}
\int d^Dl_1\,d^Dl_2\,d^Dl_3\, \frac{1}{l_1^2\, (p+l_2)^2\, l_3^2\, \dots l_8^2}\, ,
\end{eqnarray}
where $l_4 = l_3-q$, $l_5 = l_1-l_2+l_3-q$, $l_6 = l_1-q$,
$l_7 = l_2-l_1$ and $l_8 = l_2-l_3$, see ref.~\cite{Larin:1991fz} 
for the conventions of the momentum flow.
The result for the $N$-th Mellin moment of this diagram is given 
by the coefficient $c_N$, 
\begin{center}
\begin{picture}(105,30)(15,10)
\SetScale{0.5}
\SetWidth{1.2}
       \Line(35,70)(85,70) \Line(35,20)(85,20) 
       \Line(35,70)(85,20) \Line(85,70)(35,20) 
        \CArc(35,45)(25,90,270) \CArc(85,45)(25,270,90)
        \Line(10,45)(0,45) \Line(110,45)(120,45)
        {\SetColor{Red} \SetWidth{4} 
          \Line(35,70)(85,70)} 
\SetColor{Black}
        \PText(60,80)(0)[l]{1} 
        \PText(60,13)(0)[l]{1} 
        \PText(40,45)(0)[l]{1} \PText(80,45)(0)[r]{1}
        \PText(15,70)(0)[lb]{1}
        \PText(105,70)(0)[rb]{1}
        \PText(15,20)(0)[l]{1} \PText(105,20)(0)[r]{1}
        \put(65,22){\parbox{3.5cm}{
\begin{eqnarray}
= c_N \, \left(\frac{p \cdot q}{q^2}\right)^N\, .
~~~~~~~~~~~~~~~~~~~~~~~~~~~~~~~~
          \label{no2bbbpic}
\end{eqnarray} 
}}
\end{picture}
\end{center}
The solution for $c_N$ is expressed in terms of harmonic sums of weight six, 
as it is expected for the finite terms of a three-loop diagram. 
We obtain for $c_N$,
\begin{eqnarray}
%
{\lefteqn{
c_N = }}
  \label{no22bbbres} \\
&&
  \frac{(-1)^{N}}{N+1} \Bigl(
         8 S_{-3,-2}(N\!+\!1)
       + 8 S_{2}(N\!+\!1) \zeta_3 
\nonumber\\
&&
       + 4 S_{2,-3}(N\!+\!1) 
       - 4 S_{2,3}(N\!+\!1) 
\nonumber\\
&&
       + 4 S_{3,-2}(N\!+\!1)
       + 4 S_{3,2}(N\!+\!1) 
       + 10 \zeta_5
        \Bigr)
\nonumber\\
&&
+ \frac{1}{N+1} \Bigl(
         4 S_{-3,-2}(N\!+\!1)
       + 4 S_{-3,2}(N\!+\!1) 
\nonumber\\
&&
       + 8 S_{-2}(N\!+\!1)   \zeta_3 
       + 4 S_{-2,-3}(N\!+\!1)
\nonumber\\
&&
       - 4 S_{-2,3}(N\!+\!1) 
       + 8 S_{3,-2}(N\!+\!1) 
       + 10 \zeta_5
        \Bigr)\, .
\nonumber
\end{eqnarray}
This illustrates nicely that our method will not only provide the 
anomalous dimensions, which are proportional to the single
pole in $\epsilon$ in dimensional regularization, but also  
the coefficient functions which are determined by the finite terms at three loops.

In a systematic study we could set up, solve and program the complete  
reduction identities for all three-loop BBB's and all genuine
three-loop CBB's of the ladder, benz and non-planar type. 
We have used FORM~\cite{Vermaseren:2000nd} for this task and performed 
checks at all stages of the calculation with the standard MINCER
routine~\cite{Larin:1991fz} by evaluating the expressions for a number
of fixed values of the Mellin moment $N$. 
Optimization of the program requires the tabulation of several 
thousand two- and three-loop integrals in order to evaluate a given 
Feynman diagram in reasonable time. 
Thus far, the creation of these tables has already used a large amount 
of computer time. 
The complete database of Feynman diagrams for the structure functions 
$F_2,F_3$ and $F_L$ has been generated with
QGRAF~\cite{Nogueira:1991ex} and contains roughly 11000 diagrams up to
three loops. 
At present, the implementation of the reduction scheme is not complete
yet. We expect that finishing the program along with the testing 
and finally the actual calculation of all Feynman diagrams still
requires a lot of work.

\section{Conclusions}

The complete NNLO perturbative QCD corrections for deep-inelastic
structure functions significantly reduce the theoretical uncertainties 
in the determination of the strong coupling $\alpha_s$ and the parton distributions.
A precise knowledge of these quantities, including a quantitative 
error estimate, will be particularly important for new hadron collider experiments.

The present approach based on the OPE, to calculate the Mellin 
moments of the three-loop structure functions seems to allow 
a successful completion. 
The approach relies on the ability to solve all nested sums as functions of $N$
in terms of harmonic sums, to set up and solve difference equations in $N$ and, 
finally, to reconstruct the complete analytical expressions of the 
results in $x$-space by means of an inverse Mellin transformation.

As far as the theoretical developments are concerned, we believe that
there is a realistic chance for very high precision measurements in
deep-inelastic scattering.


\begin{thebibliography}{10}

\bibitem{vanNeerven:1991nn}
E.~B. Zijlstra and W.~L. van Neerven,
\newblock Phys. Lett. {\bf B272}, 127 (1991); 
\newblock {\bf B273}, 476 (1991); 
\newblock {\bf B297}, 377 (1992);
\newblock Nucl. Phys. {\bf B383}, 525 (1992).

\bibitem{Moch:1999eb}
S.~Moch and J.~A.~M. Vermaseren,
\newblock Nucl. Phys. {\bf B573}, 853 (2000), hep-ph/9912355;
\newblock Nucl. Phys. Proc. Suppl. {\bf 86}, 78 (2000), hep-ph/9909269;
\newblock {\bf 89}, 137 (2000), hep-ph/0006053.

\bibitem{Larin:1991zw}
S.~A. Larin, F.~V. Tkachev, and J.~A.~M. Vermaseren,
\newblock Phys. Rev. Lett. {\bf 66}, 862 (1991).

\bibitem{Larin:1991tj}
S.~A. Larin and J.~A.~M. Vermaseren,
\newblock Phys. Lett. {\bf B259}, 345 (1991).

\bibitem{Larin:1997wd}
S.~A. Larin, P.~Nogueira, T.~van Ritbergen, and J.~A.~M. Vermaseren,
\newblock Nucl. Phys. {\bf B492}, 338 (1997), hep-ph/9605317.

\bibitem{Retey:2000nq}
A.~Retey and J.~A.~M. Vermaseren,
\newblock Nucl. Phys. {\bf B604}, 281 (2001), hep-ph/0007294.

\bibitem{Gracey:1994nn}
J.~A. Gracey,
\newblock Phys. Lett. {\bf B322}, 141 (1994), hep-ph/9401214; 
J.~F. Bennett and J.~A. Gracey,
\newblock Nucl. Phys. {\bf B517}, 241 (1998), hep-ph/9710364.

\bibitem{Catani:1994sq}
S.~Catani and F.~Hautmann,
\newblock Nucl. Phys. {\bf B427}, 475 (1994), hep-ph/9405388;
V.~S. Fadin and L.~N. Lipatov,
\newblock Phys. Lett. {\bf B429}, 127 (1998), hep-ph/9802290;
M.~Ciafaloni and G.~Camici,
\newblock Phys. Lett. {\bf B430}, 349 (1998), hep-ph/9803389.

\bibitem{vanNeerven:1999ca}
W.~L. van Neerven and A.~Vogt,
\newblock Nucl. Phys. {\bf B568}, 263 (2000), hep-ph/9907472;
\newblock Nucl. Phys. {\bf B588}, 345 (2000), hep-ph/0006154;
\newblock Phys. Lett. {\bf B490}, 111 (2000), hep-ph/0007362;
\newblock (2001), hep-ph/0103123.

\bibitem{Botje:1999dj}
M.~Botje,
\newblock Eur. Phys. J. {\bf C14}, 285 (2000), hep-ph/9912439;
\newblock Nucl. Phys. Proc. Suppl. {\bf 79}, 111 (1999), hep-ph/9905518.

\bibitem{Giele:1998gw}
W.~T. Giele and S.~Keller,
\newblock Phys. Rev. {\bf D58}, 094023 (1998), hep-ph/9803393;
\newblock (2001), hep-ph/0104053;
W.~T. Giele, S.~A. Keller, and D.~A. Kosower,
\newblock (2001), hep-ph/0104052.

\bibitem{Seligman:1997mc}
W.~G. Seligman {\em et~al.},
\newblock Phys. Rev. Lett. {\bf 79}, 1213 (1997).

\bibitem{Kataev:1997nc}
A.~L. Kataev, A.~V. Kotikov, G.~Parente, and A.~V. Sidorov,
\newblock Phys. Lett. {\bf B417}, 374 (1998), hep-ph/9706534;
A.~L. Kataev, G.~Parente, and A.~V. Sidorov,
\newblock Nucl. Phys. {\bf B573}, 405 (2000), hep-ph/9905310;
\newblock (2001), hep-ph/0106221.

\bibitem{Benvenuti:1989rh}
BCDMS, A.~C. Benvenuti {\em et~al.},
\newblock Phys. Lett. {\bf B223}, 485 (1989);
E665, M.~R. Adams {\em et~al.},
\newblock Phys. Rev. {\bf D54}, 3006 (1996);
ZEUS, M.~Derrick {\em et~al.},
\newblock Z. Phys. {\bf C72}, 399 (1996), hep-ex/9607002;
H1, S.~Aid {\em et~al.},
\newblock Nucl. Phys. {\bf B470}, 3 (1996), hep-ex/9603004;
H1, C.~Adloff {\em et~al.},
\newblock Eur. Phys. J. {\bf C13}, 609 (2000), hep-ex/9908059.

\bibitem{Santiago:1999pr}
J.~Santiago and F.~J. Yndurain,
\newblock Nucl. Phys. {\bf B563}, 45 (1999), hep-ph/9904344;
\newblock Nucl. Phys. Proc. Suppl. {\bf 86}, 69 (2000), hep-ph/9907387;
\newblock (2001), hep-ph/0102247.

\bibitem{Gonzalez-Arroyo:1979df}
A.~Gonzalez-Arroyo, C.~Lopez, and F.~J. Yndurain,
\newblock Nucl. Phys. {\bf B153}, 161 (1979);
A.~Gonzalez-Arroyo and C.~Lopez,
\newblock Nucl. Phys. {\bf B166}, 429 (1980);
C.~Lopez and F.~J. Yndurain,
\newblock Nucl. Phys. {\bf B183}, 157 (1981).

\bibitem{Kazakov:1988jk}
D.~I. Kazakov and A.~V. Kotikov,
\newblock Nucl. Phys. {\bf B307}, 721 (1988);
\newblock {\bf B345}, 299 (1990),
\newblock Erratum.

\bibitem{'tHooft:1972fi}
G.~'t~Hooft and M.~Veltman,
\newblock Nucl. Phys. {\bf B44}, 189 (1972).

\bibitem{Chetyrkin:1981qh}
K.~G. Chetyrkin and F.~V. Tkachev,
\newblock Nucl. Phys. {\bf B192}, 159 (1981).

\bibitem{Vermaseren:1998uu}
J.~A.~M. Vermaseren,
\newblock Int. J. Mod. Phys. {\bf A14}, 2037 (1999), hep-ph/9806280.

\bibitem{Blumlein:1998if}
J.~Bl\"umlein and S.~Kurth,
\newblock Phys. Rev. {\bf D60}, 014018 (1999), hep-ph/9810241.

\bibitem{Vermaseren:2000we}
J.~A.~M. Vermaseren and S.~Moch,
\newblock Nucl. Phys. Proc. Suppl. {\bf 89}, 131 (2000), hep-ph/0004235.

\bibitem{Borwein:1996yq}
J.~M. Borwein, D.~M. Bradley, and D.~J. Broadhurst,
\newblock (1996), hep-th/9611004.

\bibitem{Remiddi:1999ew}
E.~Remiddi and J.~A.~M. Vermaseren,
\newblock Int. J. Mod. Phys. {\bf A15}, 725 (2000), hep-ph/9905237.

\bibitem{MR2000c:11108}
A.~B. Goncharov,
\newblock Math. Res. Lett. {\bf 5}, 497 (1998).

\bibitem{Tarasov:2000sf}
O.~V. Tarasov,
\newblock Nucl. Phys. Proc. Suppl. {\bf 89}, 237 (2000), hep-ph/0102271.

\bibitem{Laporta:2000dc}
S.~Laporta,
\newblock Phys. Lett. {\bf B504}, 188 (2001), hep-ph/0102032;
\newblock Int. J. Mod. Phys. {\bf A15}, 5087 (2000), hep-ph/0102033.

\bibitem{Larin:1991fz}
S.~A. Larin, F.~V. Tkachev, and J.~A.~M. Vermaseren,
\newblock NIKHEF-H-91-18.

\bibitem{Vermaseren:2000nd}
J.~A.~M. Vermaseren,
\newblock (2000), math-ph/0010025.

\bibitem{Nogueira:1991ex}
P.~Nogueira,
\newblock J. Comput. Phys. {\bf 105}, 279 (1993).

\end{thebibliography}

\end{document}